# Addressing student models of energy loss

# in quantum tunnelling[1]

Michael C. Wittmann,[1,*] Jeffrey T. Morgan,[1] and Lei Bao[2,*]

[1] University of Maine, Orono ME 04469-5709, USA

email: wittmann@umit.maine.edu, tel: 207 − 581 − 1237

[2] The Ohio State University, Columbus OH 43210, USA

**Abstract**

We report on a multi-year, multi-institution study to investigate student reasoning about energy in the context of quantum tunnelling. We use ungraded surveys, graded examination questions, individual clinical interviews, and multiple-choice exams to build a picture of the types of responses that students typically give. We find that two descriptions of tunnelling through a square barrier are particularly common. Students often state that tunnelling particles lose energy while tunnelling. When sketching wave functions, students also show a shift in the axis of oscillation, as if the height of the axis of oscillation indicated the energy of the particle. We find inconsistencies between students' conceptual, mathematical, and graphical models of quantum tunnelling. As part of a curriculum in quantum physics, we have developed instructional materials designed to help students develop a more robust and less inconsistent picture of tunnelling, and present data suggesting that we have succeeded in doing so.

---

## I. Introduction

In a multi-year study at several institutions, we have been studying student understanding of energy in the context of quantum tunnelling.  We find that students bring common sense classical reasoning to bear when thinking about quantum tunnelling.  In every part of our study, students apply ideas as if describing a projectile passing through a physical barrier and losing (or gaining, in special cases) energy.  The response that "particle energy is lost in tunnelling" is prevalent across all our studies.

Quantum tunnelling is an excellent culminating activity for students learning quantum physics.  To correctly understand tunnelling, students must understand wave functions and probability densities, potential energy diagrams, the quantum continuity conditions imposed by the Schrödinger equation, and the difference between static and stationary states.  For instructors, tunnelling offers chances to pull together different themes of a quantum physics course, to study the coherence of student understanding, and to lay the groundwork for future studies in both theoretical and applied quantum physics.

Previous research, reported in this journal, has discussed the interpretation of probabilities in the context of tunnelling [1].  Our paper complements the previous paper because it emphasizes energy and not probability.  We also use a slightly different methodological approach to investigate student reasoning, and include a discussion of a different type of instruction.

In this paper, we will present data of student reasoning about quantum tunnelling, gathered in several studies at two institutions.  We then describe an intervention and a comparison between traditional and modified instruction.  Our data are only suggestive, but indicate that our instruction is effective in addressing students' conceptions of energy loss in tunnelling.

## II.  Student understanding of tunnelling

In this section, we move from a touchstone example of a single examination question response to a more complete analysis of how this student's response is typical.

### A.  Data sources and methods of interpretation

We have gathered data from written examination questions, ungraded quizzes, specially designed surveys, and individual interviews.  For analysis, we use tools common to physics education research: content analysis of written responses, interpretation of diagrams and sketches, and descriptions of student actions.  We let the student's words speak for themselves.  Specific examples follow.  Our data come from classes taught at three levels and with many forms of instruction.  Our primary interest in evaluating the data lies in understanding the types of responses we most commonly observe.  We find that a single response, namely energy loss of particles which tunnel through square barriers, dominates in the questions that we ask.

### B.  A touchstone example of one student's detailed view of tunnelling

The final examination question shown in Figure 1 was posed to 11 University of Maryland third-year engineering students who had completed instruction on quantum physics. Students were asked to interpret typical quantities (probability, wave function, and energy) in a situation typical of a textbook problem (a particle incident on a square barrier).  Student performance on this question has been reported [2] with an emphasis on how students perform better in a modified-instruction class which used the *New Model Course in Applied Quantum Physics* [3].

We present one student's response as a touchstone example of student difficulties with the material. When asked to describe the energy of a particle that tunnelled through the barrier compared to one that had not, the student ("Dave") wrote, "[the particles] in region 3 'lost' energy while tunnelling through the barrier." Seven of the 11 students gave a similar response. The phrasing occasionally varied; another student stated, "[the particle] collides and loses energy in the barrier."

Dave's sketch of the particles' wave function is shown in Figure 2. The axis around which the wave function oscillates was higher in region I (the incoming wave) than in region III. In the barrier (region II), Dave sketched an exponential decay (though both exponential increase and decay are mathematically possible and written out in mathematical notation under the function). The axis around which the wave function oscillates shifted with the exponential decay term in the barrier. Dave's wave function seems consistent with the description of energy loss, if one considers the height of the axis of oscillation to be a measure of the energy of the particle (as is often done in bound state problems). It may be that Dave's particle-based reasoning (about energy loss) guided how he sketched the wave function, though we cannot tell from the provided evidence. We find evidence of the "axis shift" in all our studies.

Dave's mathematical equations (with unspecified normalization constants) are also shown in Figure 2. Several elements are correct. In regions I and III, the wave function oscillates sinusoidally, as indicated by the imaginary exponential terms. Only one term exists for the wave function in region III, since it describes the outgoing wave. In region II, the wave function is described by exponential decay and increase. These responses are all correct, but we note inconsistencies between his mathematical, graphical, and verbal responses to the question. The oscillation axis shift for the wave function is not represented mathematically. The student describes energy loss for the tunnelling particle, but keeps the wavelength of the sketched wave function the same in regions I and III.  Our results are consistent with those from researchers studying other areas in quantum physics [4, 5].

In summary, a single response on an exam question raises several issues about how students understand quantum tunnelling.  The student believes energy is lost in the barrier, and draws an "axis shift" in the wave function.  However, his mathematical description is inconsistent with his "axis shift" response.  At least in preliminary results, we also find that Dave's response is not unique and is common in his class.  To understand such a response in more detail, we next describe results from individual student interviews and a survey used to investigate student reasoning about quantum tunnelling.  We present data on examination data below, after describing a curriculum which affects student reasoning.

### C. Interviews to understand tunnelling through a barrier

We have used individual demonstration interviews with students both before and after they study quantum tunnelling.  Ten interviews with second and fourth year students at the University of Maine were transcribed, analyzed with regard to the conceptual models students used, and annotated according to gestures and important sketches drawn by students.

Results were consistent with the touchstone example described above.  Students often spoke of the energy loss of quantum objects that are found to the right of a square potential barrier after being incident from the left.  Common sketches, shown in Figure 3, also show the "axis shift" response.  When describing figure 3(b), "Selena" stated that "it requires energy to go through this barrier," and, "There's a possibility it will go through even though it doesn't have enough kinetic energy to *overcome* the potential energy barrier, but it still may make it through to the other side" (emphasis added).  Selena used both energy and wave function to label the vertical axis. Her confusion of energy, wave function, and probability is discussed in more detail in [6].

Selena was not the only student to give this response, even though it was not explicitly asked for. Instead, many students provided this information on their own in response to other, more general questions. For example, in a post-instruction interview (the student had completed modern physics, but not quantum physics):

Interviewer: Is there any chance the electron will ever be found in Region [III]?

"Jack": Yes, there is.

I: There is? How do you know that?

J: Well I know because I was taught that [ ... ] when the particle of some certain potential energy, or of some energy, encounters a potential barrier, there is a possibility, calculated through, well, wave equations and their integrals, that a particle will actually just go straight on through, *losing energy as it does so*, and come out on the other side of the potential barrier at a lower energy and continue on its path. (emphasis added)

We see that the energy loss and "axis shift" responses are given by many students who use a language that is remarkably consistent from student to student. Not surprisingly, when physicists' common terminology includes "barrier" and "tunnel through," we find that some students think of physical, macroscopic tunnels when they reason about quantum mechanical tunnelling. Selena's use of "overcoming" has been noted. Another interview subject discussed her mental picture of snowballs flying through snow banks. Others, in classroom settings, have spoken of bullets shot through wood planks. On the whole, we find that students rarely make *explicit* use of such analogies, but find that they often reason *as if* making such analogies.

**D.** **Building a survey to investigate student use of energy loss in tunnelling**

We now illustrate that the "axis shift" and energy loss responses are common on specially designed surveys, as well. To evaluate how common the responses in the previous section are among students studying quantum physics, we developed a short diagnostic survey on basic concepts of quantum tunnelling [7]. The questions are similar to those asked in Figure 1 in that they contain a square barrier and particles incident from the left with an energy less than the barrier's. Variations on the question included changing the width or height of the barrier and changing the energy of the incoming particles. Thirty-four students have answered two versions of the survey. (The second version is a slight modification of the first based on early responses.)

The survey starts by asking students to describe the energy of particles transmitted through a square barrier (similar to Figure 1, part c). Regions of space are defined as before, with particles incident from Region I and tunnelled particles in Region III. Results are shown in Table 1 (note that not all students answered this particular question). The most common response was to describe particle energy loss. Students were prompted to explain the reasoning used to determine their response. Answers were consistent with the interview results and the touchstone example, and included:

- "Some energy is dissipated as the particle tunnels through the potential barrier"

- "It will take some energy for the particles to penetrate the barrier in Region II"

- "Energy is 'lost' getting through the barrier"

- "The potential barrier Region II lessens the energy of the particles"

- "Particle should lose energy tunnelling through a barrier"

A subset of respondents (17 of 34) were asked to sketch the wave function for this situation. The question was nearly identical to that in Figure 1, part a. All of the students who gave the axis shift response also described energy loss. These results are consistent with those from the examination question, described previously. Their sketches are interchangeable with those in Figure 3.

From the survey, we see further evidence that the basic results of the touchstone example (energy loss while tunnelling and a sketch including an axis shift) are common to students who have already completed instruction on quantum tunnelling. Examination results, given below, give further examples of these two results.

### E. Simplifying the situation to a single step

We recognize that particles tunnelling through a barrier might be leading students to answer these questions in a macroscopic sense. To students, a barrier may cause particles to lose energy as they tunnel, but the unique challenges of transmission and reflection at a potential step may elicit a different class of responses. To understand student reasoning about tunnelling at a potential step might give us more insight into student thinking about the quantum elements of tunnelling, unburdened by classical imagery and thoughts of "overcoming" a barrier.

We studied tunnelling at a step [8] by conducting interviews with students (N = 11) and posing multiple-choice examination questions to students (N=24) at the University of Maryland. The questions deal with particles incident on step-up and step-down potentials. A typical schematic for such a situation is shown in Figure 4. The energy of particles incident on the step is always higher than the potential step (*i.e.*, E > 0 in all studied situations).

We found conflicting results on whether or not difficulties similar to those found with the square barrier occurred in such situations. Again, macroscopic descriptions dominated student thinking about the problem. When moving from a lower to a higher potential (on the right, in Figure 4), some students correctly stated that particles might be reflected, but not because of wave reflection at a step. For example, one student said "the reflectance (*sic*) is greater [for a step-up potential] due to the fact that it is kind of running into a wall in which you have more reflectance." Seven of 11 interview students stated that the reflection coefficient of the step-down system is either zero or less than the coefficient for the step-up potential. (A correct answer is that they are equal.) Reasoning commonly used to describe the lower reflection (and higher transmission) coefficient for the step-down potential was that the step-down potential "could be modelled as a ball going down and suddenly getting a whole lot of energy and it will speed up and the transmittance is getting larger." Another student said, more simply, "the particle should be attracted more to the lower potential" in region II of the step-down potential. Similar difficulties in understanding the connection between potential energy diagrams and probabilities are discussed in [9]. On a similar final examination question (with no interviewed students answering the question), 11of 24 students gave such responses.

We find that students have some of the same issues when dealing with steps as they do when dealing with barriers. Though the idea of "overcoming" a step is not as prevalent, students think of attraction to low potential energy locations and describe steps as walls an object runs into. Again, many use classical intuitions to think about quantum systems.

## III. Curriculum to address students' macroscopic models

Much of the research in this paper was conducted as part of a project to develop the *New Model Course in Applied Quantum Physics* [3], a series of small-group guided inquiry worksheets designed to complement lecture instruction for a modern physics or introductory quantum physics course. Materials have been designed similar to the *Tutorials in Introductory Physics* [10]. The full set of quantum physics worksheets is now available as *Activity-Based Tutorials Volume 2: Modern Physics* [11]. We give details of the tunnelling tutorial and present suggestive evidence of its effectiveness in comparison to more traditional, lecture-only instruction.

## A. Instructional materials

The tutorial is designed to allow comparison between classical physics and quantum systems. In the non-computer version of the tutorial (provided in the Appendix), students start by imagining a bead with charge $q$ on a wire with a single insulating spacer (see top of Figure 5a). To the right of the spacer, there is a constant potential, $V_0$. Note that we do not tell the students about the physical system which creates such a potential (it would require treating the insulating spacer as a capacitor, for example, and we wish to have students focus on other physics in this situation). Students are asked to sketch a potential energy diagram. There is a potential labelled on the wire, so the potential energy of the bead is $qV_0$ to the right of the spacer. Students describe the motion of the bead when its initial energy is $E > qV_0$ or $E < qV_0$. The next activity asks students to sketch the potential energy diagram and describe the motion of the bead when there is another insulating spacer (see Figure 5b). Students are again asked to describe the motion of a bead with different initial energies. In both the step and barrier activities, we permit students to ignore issues related to accelerating and decelerating charges. Also, we are satisfied if students approximate and draw square steps and barriers. Figure 5 shows the types of potential energy diagrams we expect students to draw in each situation.

Students then move to quantum systems, and are given the potential energy diagrams of potential steps and barriers shown at the bottom of Figures 5a and 5b. Qualitatively, these should look exactly like the student diagrams for the bead on a wire. Students are told that "a quantum particle with energy $E < U_0$ [is] incident from the left on the potential step [or barrier]." They are asked to "Describe the general shape of the incident wave function" in each region.

Based on previous activities in which students have analyzed wave functions based on whether the energy of the quantum object is greater or less than the potential of the system at that point, we expect students to state that regions I, III, and V have sinusoidal wave functions and regions II and IV have exponential wave functions. Note that students are *not* asked to draw the actual wave functions, only to describe them. The tutorial ends with a discussion of the classical limit and questions about different behaviours for more massive particles, such as protons rather than electrons. In such situations, we expect the amplitude of the sinusoidal terms in region V to be very, very small.

Students working through the tutorial give responses consistent with the results from our basic research. They do many things well. Students are usually able to draw the correct potential energy diagrams, though some need assistance to recall $U = qV$. They do not have problems with boundary conditions of wave functions at the edges of the potentials, nor are they particularly bothered by the non-zero probability of finding a particle inside a potential step. Surprisingly, in a tutorial where they are not asked to draw wave functions, we find that most students do so. Unfortunately, many draw the axis shift response. During classroom facilitation, this allows an instructor to ask questions about what the axis represents, whether the students' responses are consistent with the students' knowledge of graphs, and so on. A successful response (when drawing the wave function) requires that students find coherence between the wave nature and the particle nature of the quantum elements. Due to the length of class periods, students rarely get to the final section on the classical limit.

### *B.* Evidence of student learning

We carried out a comparison study at the University of Maryland. Two courses were taught, one with three lectures a week, and one with the tunnelling tutorial replacing one of the three lectures. The total amount of time spent on the topic was equal. Furthermore, the instructor of the lecture-only course was fully aware of the common student difficulties with energy loss in quantum tunnelling. Eleven students were taught in the lecture-only course, and thirteen were taught in the tutorial class. The question shown in Figure 1 was written by both instructors, each contributing elements and together determining what correct answer was expected in each situation.

In comparing the two courses, we emphasize several points. First, the courses were not taught at the same time. One was taught in the fall semester and one in the spring semester. (Dave, described earlier, took the fall semester course.) We have found that other courses show differences in student performance in such a situation, but were unable to control these variables for this study. Second, in neither course was the axis shift response specifically addressed before instruction. Both instructors drew wave functions at times, but were not aware of the common axis shift response. Third, the final examination question shown in Figure 1 was not seen by the tutorial students, though it had been given in the previous semester. Final examinations from the fall semester were not returned to students, so the question was unknown to the spring students. Finally, and most importantly, there were so few students in each course that the data are only suggestive, not conclusive. We recognize the need for follow-up studies.

The data come in three parts. Recall that Dave's responses included acceptable mathematical functions, a written description of energy loss, the graphical axis shift response, and describing the wave function in each region of space. Graphs showing student performance on mathematical functions, energy loss, and graphing are shown in Figures 5, 6, and 7, respectively. Each chart contains information about both the traditional (lecture-only) ($N = 11$) and modified (tutorial instruction) ($N = 13$) classes.

Both populations showed great success in writing out appropriate mathematical equations (see Figure 6). Coefficients were undetermined, and both responses with imaginary exponentials and sines and cosines were accepted.

Tutorial students did much better at describing the energy of a particle that has tunnelled through a barrier (see Figure 7). Nearly 2/3 of the lecture-only students described energy loss for a particle that has tunnelled through a barrier. This result came despite the lectures on quantum tunnelling including explicit discussion of the energy of particles that have tunnelled. Furthermore, "energy-loss" students used mathematical equations inconsistent with their conceptual understanding. The wave number, $k$, was the same in all students' equations, implying a constant energy.

The traditional (lecture-only) students were also much more likely to give the axis shift response in their examination question (see Figure 8). As described above, nearly 3/4 (7 of the 11) gave the axis shift response. It should be noted that 6 of the 7 also described energy loss in the situation and are shown as checkered data in the histogram. (The seventh "axis shift" student had given an "other" response, not easily classified, when describing energy.) Subsequent informal observations of students during instruction and instructors during faculty development workshops has strengthened our belief that the two are often observed together.

Our data suggest that the tutorial is effective in helping students deal with two major difficulties concerning quantum tunnelling. Students are less likely to assume the axis shift response (though, as noted, the tutorial never asks them to sketch the wave function). Also, students are less likely to think energy is lost when tunnelling through a barrier. However, we cannot account for which parts of the tutorial were helpful in aspect of student reasoning.

We note that subsequent versions of the tutorial have included a question in which we ask students to compare the energy of particles that have tunnelled to those which have not. We expect that such an explicit question will have a positive effect on student learning, but have not been able to do a controlled study. Subsequent instruction by other users of the instructional materials has been reported to us informally, indicating that the final results from tutorial instruction are not as good as shown here, but still better than lecture-only instruction.

## IV. Summary

In this paper, we describe two common observations. First, students often state that quantum objects which tunnel through a barrier lose energy during the tunnelling process. Second, students often sketch wave functions in which there is an axis shift in the graphical axis around which the function oscillates. These results have been observed in engineering students and physics students; in interviews, graded examinations, and ungraded surveys; with second, third, and fourth year students; and at two universities.

We have also run a pilot study to show that instructional materials can be created to help students develop a more appropriate description of the physics. The data from our study suggest that it is possible to help students think about tunnelling without thinking of losing energy in the barrier, and that it is possible to help students draw graphs more appropriately. We have reports of instruction at other institutions which supports our results.

Quantum tunnelling is notoriously difficult to understand and to teach. Students have major difficulties understanding some of the quantities involved, including potential energy diagrams, the wave function, and the meaning of probability. We observe that students' responses can be used *as if* they are thinking of macroscopic objects passing through barriers. Further research is being carried out at the University of Maine to investigate these results.

## V. Acknowledgments

The work done at the University of Maryland occurred while M.C. Wittmann and L. Bao were graduate students working with Edward F. Redish. We thank him for his many contributions to the content of the research and the paper. We thank the unnamed instructor who allowed us to carry out a study in his lecture-only course. We also thank Rachel E. Scherr, John R. Thompson, Eleanor C. Sayre, and an unnamed reviewer for their many contributions in preparing this manuscript. The work described in this paper was funded in part by National Science Foundation grants DUE945-5561, DUE965-2877, and DUE0410895, and FIPSE grant P116B970186.

# VII. Figures

## 1. *Figure 1*

Consider a beam of electrons with energy $E_0$ incident from the left (x < 0) on a potential barrier of height $U$ and width $a$ (see energy diagram to the right). Three regions are indicated on the energy diagram as *I*, *II*, and *III*.

a. Sketch the shape of the wave function of an electron in regions *I*, *II*, and *III* in the diagram to the right. Explain how you arrived at your answer.

b. Write equations for the wave function in each of the regions in the diagram. Leave normalization constants unspecified.

c. Compare the energy of electrons found in regions *I* and *III*. Explain how you arrived at your answer.

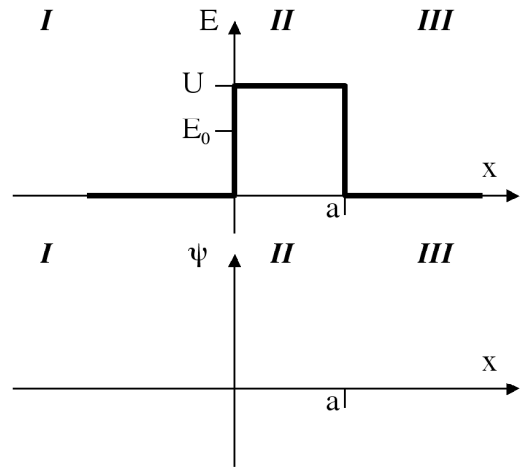

Figure 1: Final examination quantum tunneling question

## 2. *Figure 2*

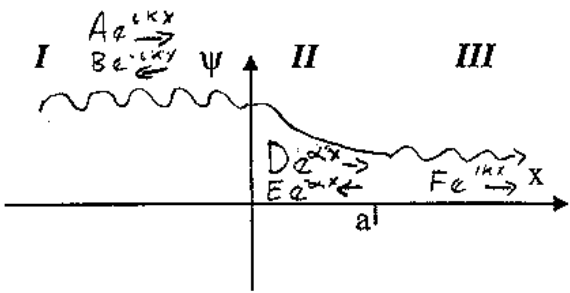

Figure 2: Dave's sketch and mathematical equations

3. *Figure 3*

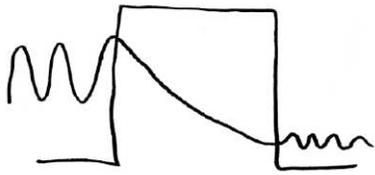

Figure 3(a): Student sketch with axis shift

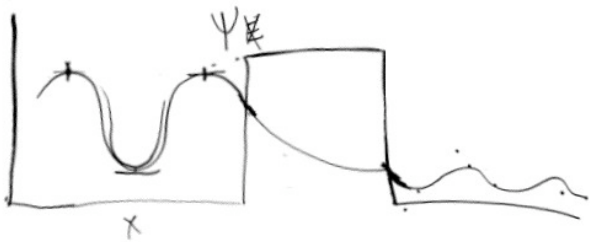

Figure 3(b): Axis shift response with multiply labeled graph axes (Selena's response)

4. *Figure 4*

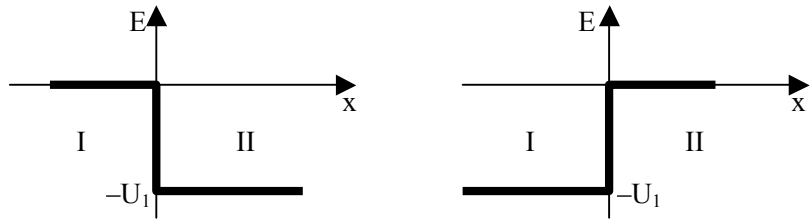

Figure 4: Step-down and step-up potential energy diagrams

## 5. *Figure 5*

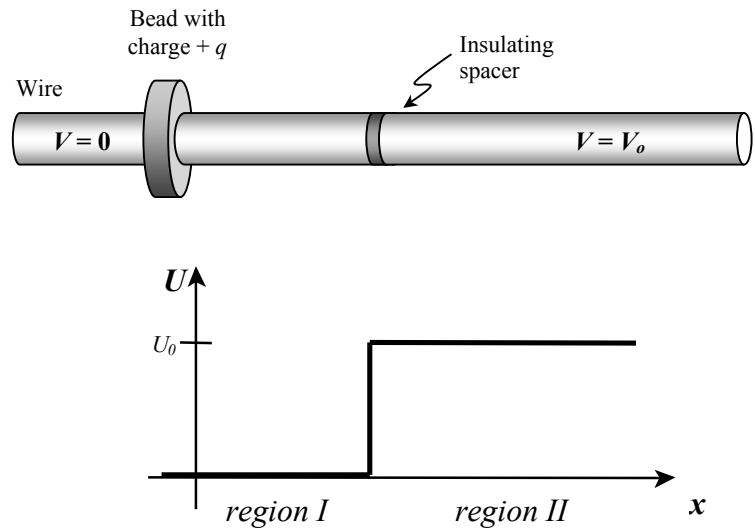

Figure 5a: Tutorial potential step activity. Students describe
a charged bead and charged wire and a quantum mechanical step.

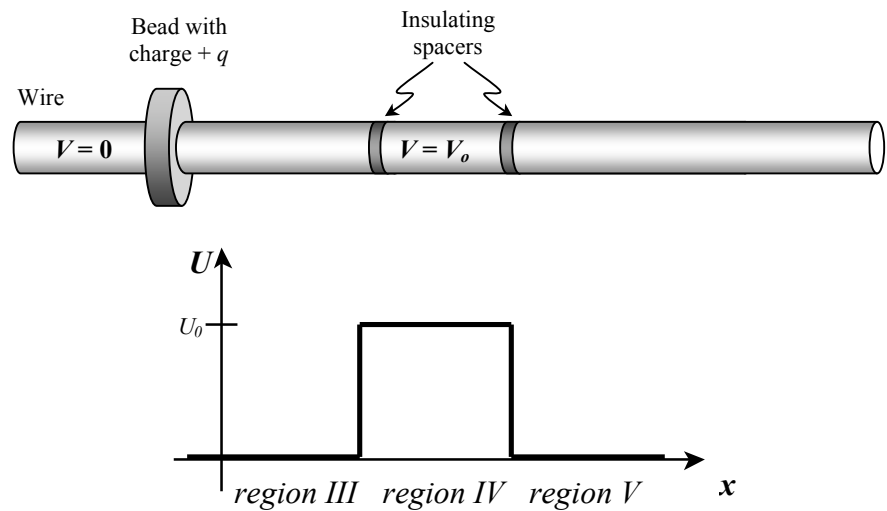

Figure 5b: Tutorial potential barrier activity. Students describe
a charged bead and wire and a quantum mechanical barrier.

### 6. *Figure 6*

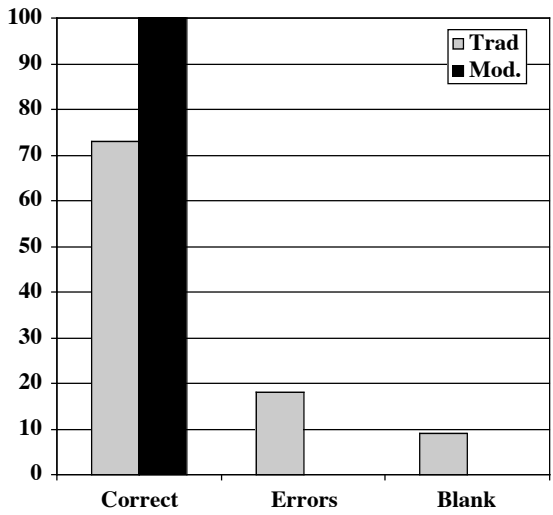

Figure 6:  Students' mathematical equations describing wave functions in three regions of space

### 7. *Figure 7*

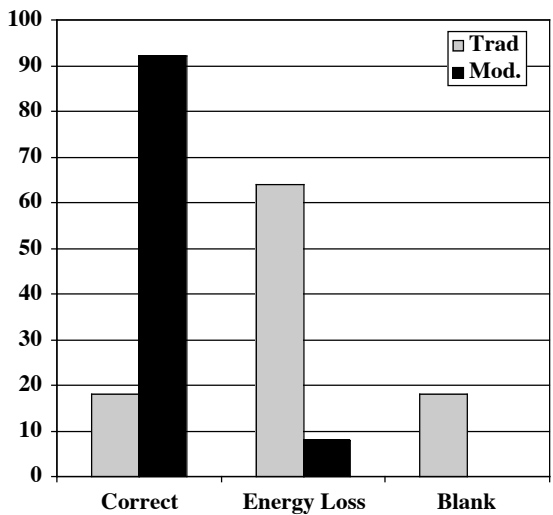

Figure 7:  Student descriptions of tunneled particle energy

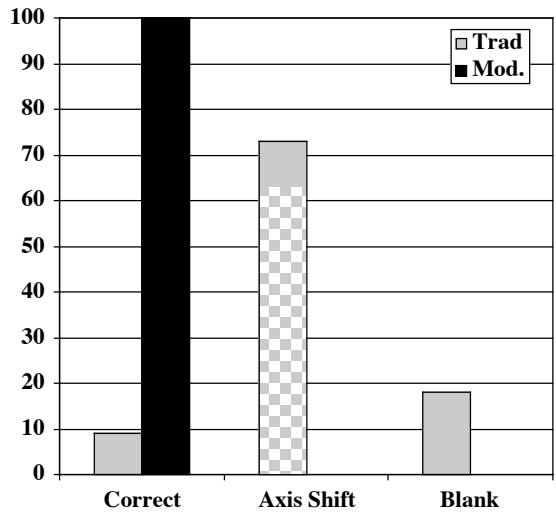

Figure 8: Student sketches of the wave function for particles incident on a square barrier (Checkered area indicates students who also gave the energy loss response)

## VIII.   Tables

### *1. Table 1*

| <u>Response</u> | <u>percentage</u><br>(*n* = 34) |
|---|---|
| *Energy in Region III is the same as the energy in Region I (correct)* | 33%<br>(n=11) |
| *Energy in Region III is less than the energy in Region I* | 64%<br>(n=21) |

Table 1: Energy of particles tunnelling through a square barrier (as in Figure 1)

**Appendix:**
***New Model Course in Applied Quantum Physics***
**Tutorial on Quantum Tunneling**

# Tunneling

## I. Classical Mechanics

### A. Potential Step

Consider a bead with a positive charge q initially moving to the right along a smooth wire. (See figure below.) The left half of the wire is grounded (V = 0). The right half of the wire is kept at a potential V = $V_0$. The two halves are separated by a small insulating spacer. (Assume that the bead is electrically insulated from the wire and ignore friction.)

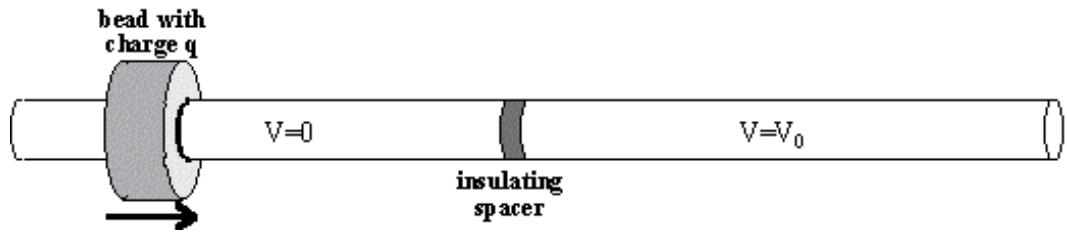

1.  In the space below, sketch a graph of the potential energy U of the bead as a function of horizontal position.

2.  Describe the entire motion of the bead if its initial kinetic energy E is such that:

    E < $qV_0$

    E > $qV_0$

### B. Potential Barrier

The experiment from part A is repeated except the wire is divided into three parts, as shown in the figure below. Again the bead has charge q and is incident from the left.

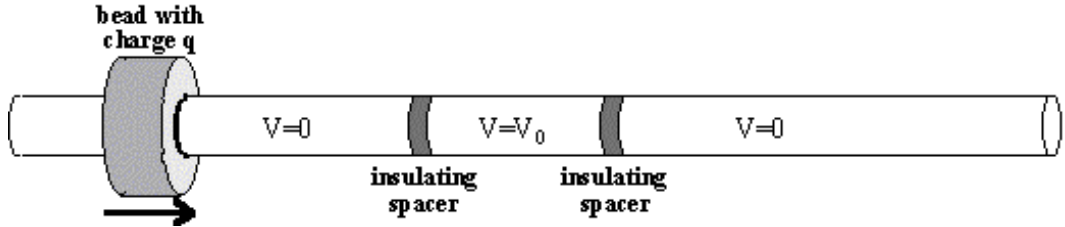

1. In the space below, sketch a graph of the potential energy U of the bead as a function of horizontal position.

2. Describe the entire motion of the bead if its initial kinetic energy E is such that:

    $E < qV_0$

    $E > qV_0$

## II. Quantum Mechanics

### A. Potential Step

Consider a quantum particle with energy $E < U_0$ incident from the left on the *potential step* sketched at right.

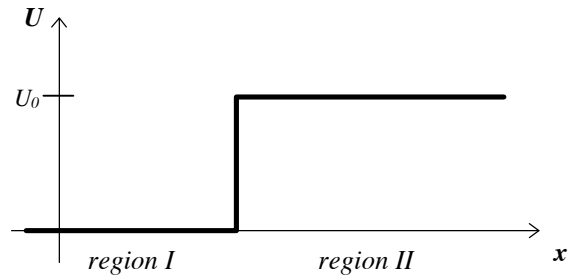

1. Describe the general shape of the incident wave function in region I. Explain how you know.

    Describe the general shape of the wave function in region II. Explain how you know.

2.  Is it possible to find the particle in a region where E < U? Explain how you know. Compare this with the classical motion.

### B. Potential Barrier

Now consider a quantum particle with energy E < $U_0$ incident from the left on the *potential barrier* as shown in the figure at right.

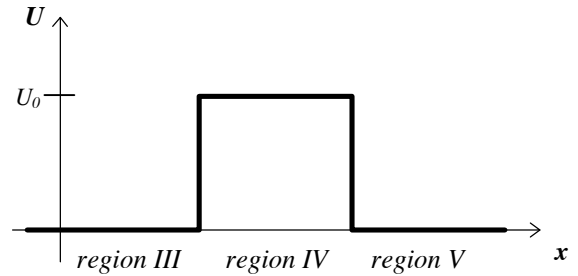

1.  Describe the general shape of the incident wave function in region III. Explain how you know.

    Describe the general shape of the wave function in region IV. Explain how you know.

    Describe the general shape of the wave function in region V. Explain how you know.

2.  It is often stated that a particle can quantum mechanically *tunnel* through a barrier. Explain what is meant by this.

3.  Consider a particle that has tunneled through the barrier. Was the energy of a particle in region III *greater than*, *less than*, or *equal to* the energy of the particle in region V. Explain how you arrived at your answer.

    Is the answer you gave consistent with the wave functions you described in question 1? Resolve any discrepancies.

## C. Classical Limit

Suppose a particle of the same energy, but greater mass, is incident on the barrier of part B.

1.  How would the shape of the wave function in each of the three regions change?  Explain your reasoning.

    Compare the probability densities in each region for part B with the probability densities in each region for the case of a more massive particle.  Explain your reasoning.

2.  Are the quantum results for very massive particles consistent with classical mechanics?  If not, resolve any discrepancies.

## D. Particle Beam

Consider a particle with kinetic energy E incident on a barrier of height $U_0 > E$.  A detector is placed on the far side of the barrier.

1.  Describe the possible outcomes of this experiment.

2.  Consider a current of N particles per second, all with kinetic energy E, incident on the same barrier.  Describe the possible outcomes of this experiment.